\documentclass[runningheads]{llncs}

\usepackage{graphicx}
\usepackage{xspace}
\usepackage{xcolor}
\usepackage{acronym}
\usepackage{enumitem}
\usepackage{amsmath}
\usepackage{adjustbox}
\usepackage{hyperref}
\usepackage{soul}

\newcommand{\pep}{$p\equiv p$\xspace}
\newcommand{\engn}{\texttt{pEpEngine}\xspace}
\newcommand{\tws}{trustwords\xspace}
\newcommand{\al}{$\mathcal{A}$\xspace}
\newcommand{\bob}{$\mathcal{B}$\xspace}
\newcommand{\eve}{$\mathcal{M}$\xspace}
\newcommand{\pepa}{$\mathtt{pEp_A}$\xspace}
\newcommand{\pepb}{$\mathtt{pEp_B}$\xspace}
\newcommand{\appliedpi}{\emph{applied pi calculus}\xspace}

\newcommand{\eStartH}[2]{\textit{startHandshake({#1},{#2})}}

\newcommand{\eRprS}[3]{\textit{receiver{#1}rustsS({#2},{#3})}}

\newcommand{\red}{\textsc{Mistrusted}\xspace}
\newcommand{\grey}{\textsc{Unsecure}\xspace}
\newcommand{\yell}{\textsc{Secure}\xspace}
\newcommand{\grn}{\textsc{Trusted}\xspace} 

\acrodef{eee}[E2E]{\emph{end-to-end encryption}}
\acrodef{im}[IM]{instant messaging}

\usepackage{fancyhdr}
\begin{document}
	
\title{A Formal Security Analysis of the \pep Authentication Protocol for Decentralized Key Distribution and End-to-End Encrypted Email
\thanks{This is a copy of the author preprint. The final authenticated version is available online at \url{https://doi.org/10.1007/978-3-030-39749-4_11}.}
}

\author{Itzel Vazquez Sandoval\inst{1} \and
Gabriele Lenzini\inst{1}\orcidID{0000-0001-8229-3270}}
\authorrunning{Vazquez Sandoval and Lenzini}
\institute{University of Luxembourg, Luxembourg
\email{\{itzel.vazquezsandoval,gabriele.lenzini\}@uni.lu}}

\maketitle              % typeset the header of the contribution

\begin{abstract}
	
To send encrypted emails, users typically need to create and exchange keys 
which later should be manually authenticated, for instance, by comparing long strings of characters. 
 These tasks are cumbersome for the average user.
To make more accessible  the use of encrypted email, a secure email application named \pep
automates the key management operations; \pep still requires the users to carry out the verification, 
however, the authentication process is simple: users have to compare familiar words instead of strings of random characters, 
then the application shows the users what level of trust they have achieved via colored visual indicators.
Yet, users may not execute the authentication ceremony as intended, \pep's trust rating may be wrongly assigned, or both. To learn whether \pep's trust ratings (and the corresponding visual indicators) are assigned consistently, we present a formal security analysis of
\pep's authentication ceremony. From the software implementation in C, we derive the specifications of an abstract protocol for public key distribution, encryption and trust establishment; then, we model the protocol in a variant of the applied pi calculus and later formally verify and validate specific privacy and authentication properties. 
We also discuss alternative research directions that could enrich the analysis.

\keywords{formal verification \and authentication protocols \and software security analysis \and privacy-by-default \and secure email \and end-to-end encryption}
\end{abstract}

\section{Introduction}
\label{sec:intro}
Despite the success of \ac{im} applications, email prevails as the principal means for written communication \cite{RadicatiGroup2018}; yet, communication over email remains largely insecure nowadays \cite{clark2018securing}.
Solutions for securing email have however been proposed. For instance, 
OpenPGP \cite{openPGP} is arguably the most widely used email encryption standard. Derived from the PGP software, it proposes the use of symmetric and asymmetric cryptography plus data compression to encrypt communication, and digital signatures for message authentication and integrity. 

Unfortunately, severe usability drawbacks have been identified and highlighted in the standard (e.g. \cite{whitten1999johnny}).
Along with the need for users to understand at least general cryptographic concepts regarding encryption---which inevitably narrows down the scope of the audience---the principal issue is the need for verifying the ownership of public keys, i.e., that a public key claimed to be of an entity $A$ does indeed belong to $A$ exclusively. Various approaches tackle this problem, e.g., fingerprint comparisons, public key infrastructure, certificate authorities, and the notion of web of trust, which involves individuals signing each other's public keys, thus forming a chain of certifications \cite{zimmermann1995official}. However, these approaches have encountered limited adoption mostly due to usability or scalability issues \cite{clark2018securing}.

Attempting to overcome OpenPGP's usability issues related to trust establishment,
an open source commercial software, called \pep (Sec. \ref{sec:pep}), proposes the use of so called \emph{\tws} (detailed in Sec. \ref{sec:trustwords}) to carry out peer-to-peer entity authentication via an out-of-band channel---e.g., in-person, video-call. This approach argues to introduce an improvement to usability and security of the PGP word list.

In this work we present a formal security analysis of the core protocols implemented in \pep, focusing particularly in authentication and privacy goals.

\subsection{Contributions}
First, we derive from the open source code the specifications of \pep's abstract protocols for key distribution and trust establishment, and present them as Message Sequence Charts (MSC). From now on, we will refer to this abstraction as the \pep \emph{protocol}.
This is the first detailed technical documentation of such protocol.

Second, we provide a symbolic formal security analysis of the \pep protocol with respect to authentication and privacy goals, under a Dolev-Yao threat model. The analysis validates the security claims of \pep and the correct assignment of privacy ratings to messages.

\section{Context and Approach}
\label{sec:context}
The application of formal methods for verifying that specific security properties hold in cryptographic protocols in the presence of a certain adversary is a well-established research area.
Both the detection of flaws in a protocol (or, contrariwise, the proof of security) and the nature of those flaws depend on different factors, such as the verification approach and the phase of the system in which it takes place (e.g., design, implementation, compilation).
An introductory reference for the topic is \cite{sjouke:2012}.

A variety of tools and formalizations have been used to successfully analyze, amongst others, authentication scenarios in real world and authentication standards (e.g., \cite{basin2013provably,basin2018formal,cremers2011key}). Important flaws have been discovered even in well-established protocols years after their publication and while being used (e.g., \cite{loweAttack}). 
Therefore and because the design of protocols is by default an error-prone task, to effectively protect a system, security protocols need to be not only carefully designed and rigorously implemented but also strictly verified. %for flaws

Here, we carry out a symbolic formal analysis of the \pep protocol specification.
The symbolic approach assumes cryptographic primitives to work as perfect black boxes and focuses on the description of the logic of the protocol, the interaction among participants and the exchange of messages \cite{blanchet2012security}. The resulting models allow to seek for attacks that rely on logical flaws in the protocol while taking advantage of mature automated tools for protocol analysis (e.g., ProVerif \cite{proVerif}, Tamarin \cite{tamarin}).

Our work concerns remote human-to-human authentication, where human $A$ wants to be sure that human $B$ is who he claims to be and vice versa---in \pep, the owner of a specific public key---, in a global communication scenario where $A$ and $B$ might not know each other.

\subsection{Methodology}
\label{subs:methodology}
At the time when we started studying the \pep protocol there was not substantial documentation regarding neither the protocol specifications nor the source code. In consequence, the work presented here relies on the open source code of \pep \cite{pepSourceCode}, together with online documentation mainly for users \cite{pepDoc}. Recently some internet drafts have been released \cite{pepTwds,pepProtocols}, which has helped clarifying our models.

Our security analysis consists of the following steps, which we detail in the rest of the paper:

\begin{enumerate}[topsep=0pt]
	\item Extract the specifications of the key distribution and handshake protocols from the available sources (\cite{pepSourceCode,pepDoc}).
	\item Describe the protocol in MSC notation.
	\item Formalize in the \appliedpi the MSC specifications of the previous step, along with the attacker model.		
	\item Specify and formalize in the \appliedpi the properties to be verified.
	\item Verify the satisfiability of the properties formalized in 4, in the model resulting from step 3. %We use the tool ProVerif for this task.
	\item Analyze and interpret the results of the verification.
\end{enumerate}

We start by introducing the \pep software and its relevant features in Section \ref{sec:pep}.
Then, steps 1 and 2, which deal with specifying the \pep protocol, are presented in Section \ref{sec:protocols}.
In Section \ref{sec:propInformal}, we define the security properties related to privacy and authentication that concern our analysis.
Section \ref{sec:analysis} covers steps 3 and 4 of the methodology, i.e., the formalization of the protocol and of the security properties introduced informally in Section \ref{sec:propInformal}.
The results of the execution of step 5 and the analysis in step 6 are discussed in \ref{subs:results}; we also discuss limitations of the analysis in \ref{sec:discussion}.
Further directions and conclusions are presented in the last Section.

\section{Background: Pretty Easy Privacy (\pep)}
\label{sec:pep}

Pretty Easy Privacy (\pep)%\footnote{\url{https://www.pep.security}} 
\footnote{https://www.pep.security} is
a software that claims to provide privacy-by-default in email communications via end-to-end opportunistic encryption.
Roughly, this means that the software encrypts outgoing email messages without any intervention from the user, whenever 
a secure or trusted public key of the intended receiver is available.

\pep attempts to automate tasks that would otherwise require specialized-knowledge from non-expert users, while informing the user of the privacy rating assigned to messages in an intuitive way. Hence, its more relevant features are:
(1) a fully automated process for the generation and management of encryption keys and for the encryption of emails; (2) an algorithm to determine the strongest privacy level that can be assigned to a message for a specific partner---this level is further communicated to the user by colored visual icons; (3) a fully decentralized architecture for key storage---this design decision eludes relying on possibly untrusted central authorities by having the users perform the trust establishment task via out-of-band channels.

\pep is distributed as a standalone application for Android and as plugins for desktop installations of some existing email clients, e.g., Outlook, Thunderbird.
In this work we consider a general abstraction of the \pep protocols that represent improvements to PGP by means of the features described above. Comparing and discussing specific implementations is out of the scope of this paper.

\subsection{\pep Trustwords}
\label{sec:trustwords}
Manual key-fingerprint comparison is a well-established method for entity authentication in messaging protocols; yet, the approach has been shown to perform poorly for the intended goal (e.g., \cite{dechand2016empirical}).
As a solution, in addition to hexadecimal numbers, PGP allows fingerprints to appear as a series of so-called ``biometric words'', which are phonetically different English words that intend to ease the comparison for humans and to make it less prone to misunderstandings \cite{pgpWordList}.  

Trustwords in \pep follow  the same idea; they are natural language words mapping hexadecimal strings that are used to authenticate a peer after having exchanged public keys in an opportunistic manner. 
In short, such hexadecimal strings represent a combined fingerprint obtained by applying an XOR operation to the fingerprints associated to the public keys being authenticated. 
Each block of 4 hex characters of the combined fingerprint is mapped to a word in a predefined \tws dictionary.  
For instance,
\texttt{F482 E952 2F48 618B 01BC 31DC 5428 D7FA} could be mapped to \texttt{kite house brother town juice school dice broken}.

The main difference with the ``biometric words'' is the availability of \tws in different languages, which improves the security for non-English speakers, and the use of longer words, which presumably increases the entropy as the dictionary is larger and therefore the likelihood for phonetic collision is decreased \cite{pepTwds}. 
Considerations regarding the number of words in the dictionaries and the length of the words themselves are discussed also in \cite{pepTwds}.

\subsection{Trust Rating and Visual Indicators}
In agreement with the privacy-by-default principle, \pep assigns a specific privacy rating to each email exchange. Such a rating is determined per message and per identity depending on certain criteria and is shown to the users by colored icons in the message. 
% --red, gray, yellow and green, respectively--. 
The ratings are:
\begin{itemize}
	\item \red: the system has evidence that the communication partner is not who (s)he claims to be, e.g., when the user explicitly mistrusts a peer.
	\item \textsc{Unknown/Unsecure/Unreliable} (\grey): encryption/decryption of a message cannot be properly executed, e.g., when the recipient does not use any secure email solution. The message is sent in plain text.
	\item \yell: the user has a valid public key for the recipient, however it has not been personally confirmed. The message is encrypted/decrypted.
	\item \grn: the user has the recipient's public key and it has been validated with the peer. The message is encrypted/decrypted and authenticated.
\end{itemize}

\subsection{Technical Specifications of \pep}
The core component of \pep is \engn, a library developed in C99 where the automation of cryptographic functionalities (e.g., key generation) is implemented relying on existing standards and tools for secure end-to-end encrypted communications (PGP, GnuPG). 
The \pep protocols are built upon those functionalities, therefore \engn is the component from which we extracted the specifications hereby presented.

Each installation of \pep creates a local database of \pep peers, their corresponding keys and privacy ratings. Additionally, it creates a database from which the \tws for mutual authentication are retrieved; the \tws database contains the exact same data in all the distributions. To securely store private and public keys in the devices, \pep uses GnuPG\footnote{https://www.gnupg.org/}. A more detailed description of \pep can be found in \cite{pepProtocols}.

\section{The \pep Protocol}
\label{sec:protocols}
In order to carry out a security analysis it is essential to clearly understand the logic of the protocol, to know the cryptographic primitives used, the parties involved and the messages exchanged between them.
Our case study required us to obtain this information mainly from the source code of \pep. 

Following the approach in \cite{DBLP:conf/compsac/SandovalL18}, we executed the first step of the methodology proposed here in Sec. \ref{subs:methodology} by reverse engineering a fragment of the source code files. 
We then represented the output of such a process by means of MSC diagrams (step 2)
which \pep confirmed to be accurately representing their protocol. 

Here, we present and describe such diagrams which correspond to our abstracted version of the key distribution and authentication protocols used by \pep to engage in end-to-end private and authenticated communications.

In the rest of the paper, we will use $sk_x$ and $pk_x$ to refer to secret and public keys owned by agent $x$, respectively.
As well, we use \al and \bob to refer to honest participants and \eve for the malicious agent trying to prevent the honest parties from achieving the security goals.

\subsection{Public key Distribution and Encrypted Communication}
\label{subsec:keyDistProtocol}
Let \al and \bob be two partners that do not know each other's public key. 
\al installs \pep from scratch without having any cryptographic keys.
She wants to privately communicate with \bob who is already a \pep user owning a pair of keys ($sk_B$, $pk_B$). We denote the \pep instances running in \al's and \bob's devices as \pepa and \pepb respectively.

\begin{figure}[h]
	\centering
	\includegraphics[width=\textwidth]{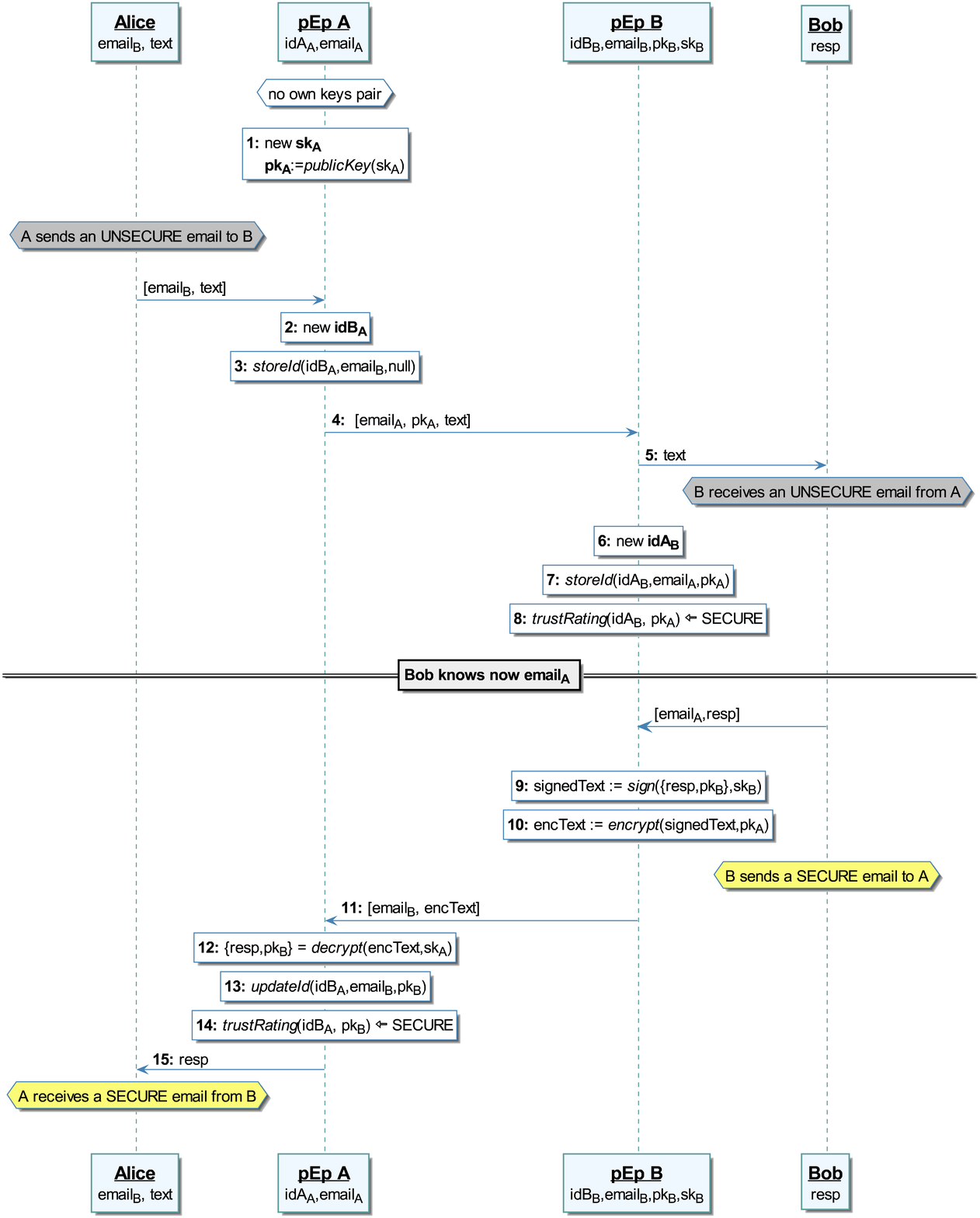}
	\caption{\pep Key Distribution Protocol}
	\label{fig:keyMngP}
\end{figure}

So that the key distribution protocol (Fig. \ref{fig:keyMngP}) can take place, when \pep is installed, \pepa generates a pair of keys ($sk_A$, $pk_A$) for \al (step 1).
The protocol starts when \al sends a message $m$ to \bob; \pepa creates an identity for \bob (2) and stores his contact details (3); then, \pepa sends $m$ as plain text along with $pk_A$ (4).
When \pepb receives the message, it displays $m$ to \bob with the privacy rating \grey (5); 
additionally, \pepb creates an identity for \al (6) and stores her email address and $pk_A$ (7); finally \pepb assigns the privacy rating \yell to \al's identity (8).
When \bob replies to \al, \pepb attaches $pk_B$ to his response $\mathit{resp}$; this message is then signed with \bob's secret key $sk_B$ (9) and encrypted using $pk_A$ (10). The signed and encrypted message is sent to \al (11); \pepb shows to \bob his message as \yell.
At reception, \pepa decrypts \bob's message using $sk_A$ (12); then it stores $pk_B$ as the public key of \bob (13) and assigns to his identity the \yell rating (14).
\bob's response is finally shown as \yell to \al.

Note that the identifiers created for \al and \bob do not need to coincide in \pepa and \pepb, since they are only used by the corresponding \pep instance. Also, $pk_A$ and $pk_B$ sent in steps (4) and (11) are only attached to the first communication between \al and \bob or whenever they are updated.

The key distribution protocol allows making the communication secret to everyone but the receiver, however, it does not guarantee that the receiver is the intended person. Man-in-the-middle attacks are still possible, as we will discuss in Section \ref{subs:results}.

\subsection{Authentication and \pep Privacy Rating Assignment}
\label{subs:handshakeProtocol}

\begin{figure}[th]
	\centering
	\includegraphics[height=\textheight]{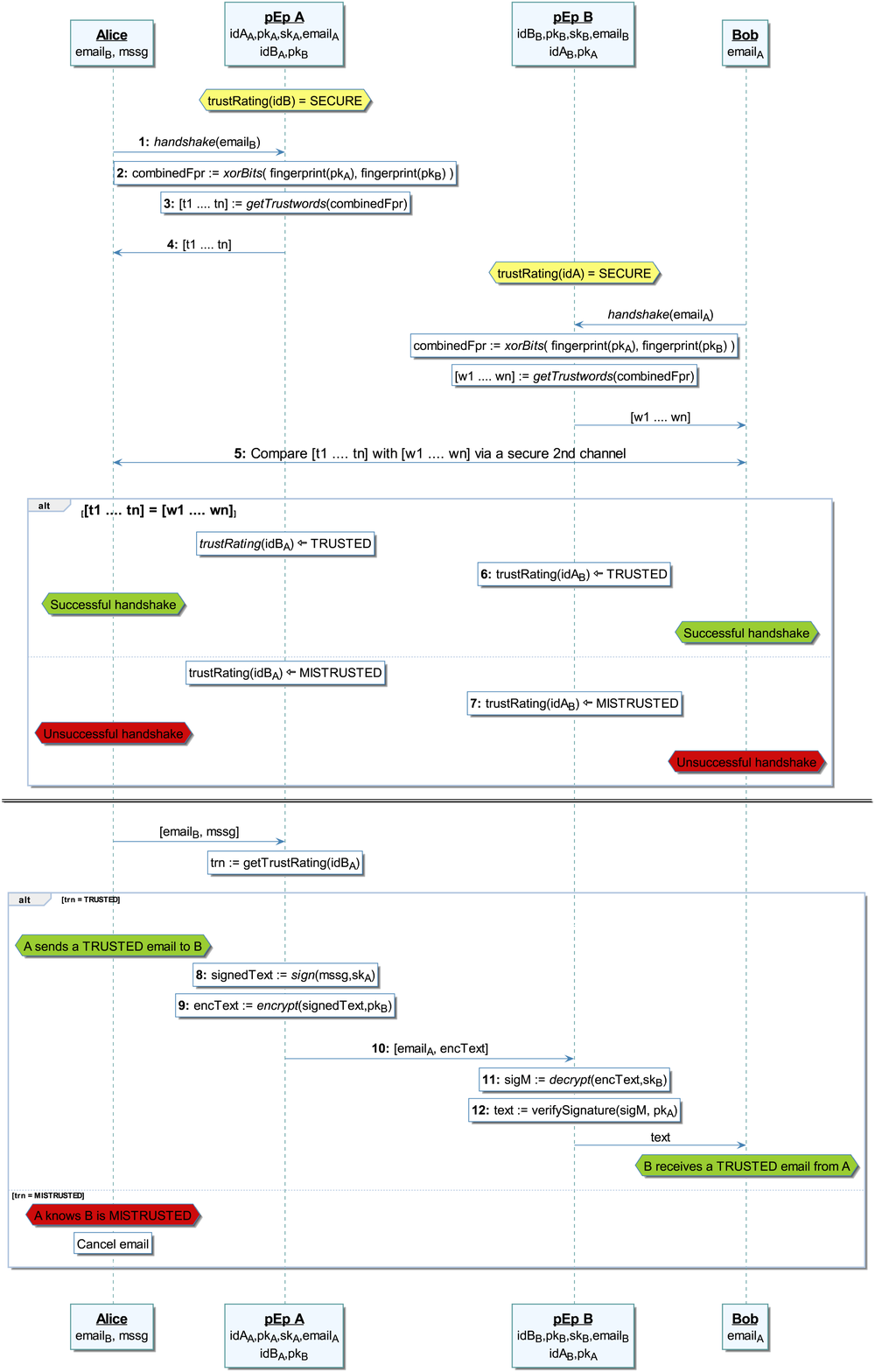}
	\caption{\pep Handshake Protocol for authentication}
	\label{fig:hndshkP}
\end{figure}

Trust establishment is achieved via the \pep Handshake protocol (Fig. \ref{fig:hndshkP}), which consists in \al and \bob comparing a list of \tws via a communication channel assumed to be secure and that needs to be used only once.

When \al selects the option to perform a handshake with \bob (1), \pepa generates a combined fingerprint based on applying an $xor$ function to the fingerprints of \al and \bob (2). The resulting hexadecimal string is mapped onto words in the selected language from the \tws database (3) and displayed to \al (4).
The analogous actions occur in \pepb when \bob selects the handshake option.
Given that the \tws database is the same in all \pep distributions, if \pepa and \pepb use the same input parameters, i.e., the same public keys and thus the same fingerprints, the list of \tws generated by each \pep instance must be the same.

The next step is the authentication, where \al and \bob contact each other in a way that they are sure to be talking with the real person, and compare the list of \tws displayed for each (5). 
If \bob confirms that the list of \tws given by \al matches exactly the one shown in his device, \al's privacy rating is set to \grn (6); we call this case a \emph{successful handshake}. Conversely, in an \emph{unsuccessful handshake} \al's rating is downgraded from \yell to \red (7).
The analogous occurs in \al's device with respect to \bob.
The privacy rating assigned after a handshake remains for all future exchanges with the communication partner.

After a successful handshake, the communication between the identities %of \al and \bob
that performed the handshake is always encrypted and authenticated (8-12).

Remark that \pep does not force users to perform the handshake protocol. The email messages are always sent regardless of the security level, which is decided per message and per recipient according to the recipient's data available.

\section{Security Properties}
\label{sec:propInformal}
Our requirements for authentication match the definition of \emph{full agreement} given by Lowe in \cite{lowe1997hierarchy}. This definition subsumes aliveness, weak agreement, non-injective agreement and injective agreement as defined in the same reference; broadly, it requires the two participants to agree on all the essential data involved in the protocol run, in our case, the public keys $pk_A$ and $pk_B$ and the email addresses.

\begin{definition}[Full agreement, from \cite{lowe1997hierarchy}]
	A protocol guarantees to an initiator A full agreement with a responder B on a set of data items ds if, whenever A completes a run of the protocol, apparently with responder B, then B has previously been running the protocol, apparently with A, and B was acting as responder in his run, and the two agents agreed on the data values corresponding
	to all the terms in ds, and each such run of A corresponds to a unique run of B. Additionally, $ds$ contains all the atomic data items used in the protocol run.
\end{definition}

Here we redefine this property in terms of \pep and introduce informally other properties in which we are interested. 

\begin{property}[Full agreement]
	\label{p:fullAg}
	A full agreement between \al and \bob holds on $pk_A$, $pk_B$, $email_A$ and $email_B$ if, whenever \al completes a successful handshake with \bob, then: \bob has previously been running the protocol with \al, the identity data of \al is ($email_A$, $pk_A$) and the identity data of \bob is ($email_B$, $pk_B$).
	%Full agreement on $pk_A$, $pk_B$, $email_A$ and $email_B$ holds between \al and \bob, if previously: \al started a handshake with \bob, \bob started a handshake with \al, the identity data of \al is ($email_A$, $pk_A$) and the identity data of \bob is ($email_B$, $pk_B$).
\end{property}
Recall that a successful handshake is only reached if \bob confirms that the \tws given by \al match exactly those shown in his device, and vice versa; therefore, the agreement on the \tws is implicit in the definition.

\begin{property}[Trust-by-Handshake] 
	\label{p:trustByH}
	Trust-by-Handshake holds for \bob if whenever \bob receives a message with privacy rating \grn  from \al, then previously \bob executed a successful handhsake with \al.
\end{property}

\begin{property} [Privacy-from-trusted] 
	\label{p:pFt}
	Privacy-from-trusted holds for \bob if, whenever \bob receives a message $m$ with a privacy rating \grn from \al, then \al sent $m$ to \bob and $m$ is encrypted with \bob's public key.
\end{property}

\begin{property}[Integrity-from-trusted] 
	\label{p:iFt}
	Integrity-from-trusted holds for \bob if, whenever \bob receives a message $m$ with a privacy rating \grn form \al, then \al sent $m$ to \bob and $m$ is signed with a valid signature of \al.
\end{property}

\begin{property}[MITM-detection] 
	\label{p:mitm}
	MITM-detection holds if whenever an unsuccessful handshake between \al and \bob occurs, then \al had previously registered a key for \bob that does not belong to him, vice versa, or both.
\end{property}

\begin{property}[Confidentiality]
	\label{p:secret}
	Confidentiality holds if \eve cannot learn the content of any message sent encrypted between \al and \bob.
\end{property}

\section{Formal Security Analysis}
\label{sec:analysis}

A security analysis requires three elements: a protocol model, a set of security properties, and a threat model defining the capabilities of the adversary by which the scope of the verification is framed.

We model the \pep protocols in the \appliedpi \cite{appliedPi}, a process calculus suitable for describing and reasoning about security protocols in the symbolic approach. Participants are represented as processes and their message exchanges are represented by terms sent over public or private channels.
A so called equational theory defines how the cryptographic operations occurring in the protocol relate with each other, and how they can be applied to obtain equivalent terms.

\subsection{Threat Model and Trust Assumptions}
\label{subsec:threatModel}
The initial assumption is that the participants have a genuine and correct distribution of the \pep software (free of implementation flaws).
To determine a relevant attacker model we need to consider the decentralized architecture of \pep.
To an attacker with access to the user's device, not only the code but also the application databases and the keys repository are available.
\eve can thus have \bob trusting her by simply modifying the corresponding record in the privacy ratings database, even if a handshake was never performed.
Modifications to the \tws database would also result in an attack, which although not threatening privacy, could prevent \al and \bob from establishing a valid trusted communication as \grn.
Therefore, we restrict the threat model with the following assumptions:
\begin{enumerate}
	\item \pep users are honest participants and their devices are secure;
	\item The adversary cannot modify exchanges over the \tws channel;
	\item The adversary has complete control over the network used to exchange emails (Dolev-Yao attacker \cite{Dolev:1981:SPK:1382435.1382728});
\end{enumerate}

These assumptions allow \eve to eavesdrop, remove, and modify emails exchanged between \al and \bob, as well as to send them messages of her choice; this includes learning their public keys exchanged by email. \eve cannot however interfere with the channel used to corroborate \tws. Remark that this is a secondary channel such as the phone or in-person, thus, not intended to replace the email communication channel.

\subsection{Modeling the \pep Protocol}
\label{subs:model}
The \pep protocol consists of the sequential execution of the key distribution and the trust establishment protocols presented in Section \ref{sec:protocols}.

\al and \bob are represented by two processes, \texttt{senderA} and \texttt{receiverB}, whose parameters symbolize the knowledge that they have. To communicate with \bob, \al needs to know his contact details, which here we abstract with the type $\mathit{userId}$;
in turn, \bob only needs to know his own id and his secret key. The actions for each participant come from the diagrams in Figures \ref{fig:keyMngP} and \ref{fig:hndshkP}. We run multiple instances of \al as well as of \bob, to simulate communication with multiple peers.

For the exchange of emails we use a public channel; on the contrary, a private channel models the \tws' validation channel.
In order to prove confidentiality of encrypted and authenticated communication, we introduce a private message $\mathit{mssg}$ representing a message whose content is initially unknown to \eve;
then, we model \al sending $\mathit{mssg}$ to \bob via the public channel after a successful handshake between them.
Since \bob is trusted,  $\mathit{mssg}$ is sent signed and encrypted (steps 8-9, Fig. \ref{fig:hndshkP}), and thus, expected to remain unreadable by \eve at the end of the protocol.

According to the symbolic model assumption, our equational theory models a perfect behavior of asymmetric encryption and digital signatures.
These equations capture the relationships allowed among the cryptographic primitives involved, determining the ways in which any participant, the attacker included, can reduce terms.
Then, for $M$ a message and $\mathit{SK}$ a secret key:

{\small
	\setlength{\abovedisplayskip}{0pt}
	\setlength{\belowdisplayskip}{0pt}
\begin{eqnarray}
	{\displaystyle \mathit{adec(aenc(M,pubKey(SK)),\ SK) = M} }\\
	\mathit{verifSign(sign(M,SK),\ pubKey(SK)) = M} \\
	\mathit{getMssg(sign(M,SK)) = M} 
\end{eqnarray}
}

Equation (1) expresses that a message $M$ encrypted with a certain public key can be decrypted with the corresponding secret key; moreover, this is the only way to obtain $M$ from a ciphertext since there is no other equation involving the $aenc$ primitive. Analogously, equation (2) returns $M$ only if it was signed with the secret key associated to the public key used for the verification. Equation (3) allows the recovery of a message without verification of a digital signature and we introduce it here to model the capability of \eve for learning messages without the need of verifying the signature.

Additionally, we assume and model that users execute the comparison of \tws correctly, i.e., they confirm the \tws in the system only when they match in the real world and they reject them only in the contrary case. This assumption implies also correctness of the trustwords generation function.
We abstract fingerprints as public keys since a PGP fingerprint is uniquely derived from a public key.
Then, for two public keys $\mathit{PK_1}$, $\mathit{PK_2}$, two \tws lists $W_1, W_2$ and the \tws generation function \emph{trustwords}: 

{\small
	\setlength{\abovedisplayskip}{0pt}
	\setlength{\belowdisplayskip}{0pt}
\begin{eqnarray*}
\mathit{trustwordsMatch(trustwords(\mathit{PK_1},\mathit{PK_2}), trustwords(\mathit{PK_1},\mathit{PK_2})) = true} \\
\mathit{trustwordsMatch(trustwords(\mathit{PK_1},\mathit{PK_2}), trustwords(\mathit{PK_2},\mathit{PK_1})) = true} \\
\mathit{otherwise} \  \mathit{trustwordsMatch(W_1,W_2) = false}.
\end{eqnarray*}
}

During its computations, \eve is allowed to apply all and only these primitives. Additionally, she has access to all the messages exchanged via the public channels and to any information declared as public. This models for instance \eve's real-life capability of generating the \tws, which is possible because all the elements are public knowledge: the source code of the function, the trustwords database, \bob's public key and \al's public key.

\subsection{Privacy and Authentication Properties of \pep}
\label{subs:formalProp}
We formalize the properties introduced in Section \ref{sec:propInformal} as correspondence and reachability queries based on events.
Correspondences have the form $E \Longrightarrow e_1 \land ... \land e_n$; they model properties expressing: \emph{if an event $E$ is executed, then events $e_1, ..., e_n$ have been previously executed}. Events mark important states reached by the protocol and do not affect the protocol's behavior.
Our properties are defined in terms of the next events, where $s$ and $r$ represent two \pep users:

\begin{itemize}
	\item \emph{endHandshakeOk(s,r,$pk_s$,$pk_r$,$e_s$,$e_r$)}: $s$ and $r$ completed a successful handshake with the public keys and emails ($pk_s$, $e_s$) and ($pk_r$, $e_r$) respectively.
	\item \eStartH{s}{r}: $s$ starts a handshake via a second-channel with $r$
	\item \emph{userKey(s,$pk_s$)}: the agent $s$ is the owner of the key $pk_s$
	\item \emph{userEmail(s,$e_s$)}: the agent $s$ owns the email address $e_s$
	\item \emph{receiveGreen(r,s,m):} $r$ received the message $m$ from $s$ as \grn
%	\item \eSprR{T}{s}{r}: the initiator $s$ sets the privacy rating of $r$ as \grn after confirming that the \tws match
	\item \eRprS{T}{r}{s}: the contacted peer $r$ sets the privacy rating of $s$ as \grn after confirming that the \tws match
	\item \emph{sendGreen(s,r,m):} $s$ sent the message $m$ to $r$ as \grn
	\item \emph{decryptionFails(r,s,m):} $r$ cannot decrypt a  message $m$ from a trusted peer $s$ 
	\item \emph{signVerifFails(r,s,m):} $r$ cannot verify the signature attached to $m$ as a valid signature of $s$
	\item \emph{endHandshakeUnsucc(s,r,$pk_s$,$pk_r$)}: $s$ and $r$ completed an unsuccessful handshake with the public keys $pk_s$ and $pk_r$ respectively.
	\item \emph{attacker($m$)}: the adversary knows the content of the message $m$
\end{itemize}

Then, for a private message $\mathit{mssg}$ and for all \pep users $a$ and $b$, messages $m$ and public keys $ka$, $kb$, $pk_A$, $pk_B$:

{\small
	\setlength{\abovedisplayskip}{0.5em}
	\setlength{\belowdisplayskip}{.5em}

	\noindent\paragraph{}
	\textbf{Full Agreement}. For email addresses $e_A$ and $e_B$,
	\begin{eqnarray*}
		\mathit{endHandshakeOk}(a,b,pk_A,pk_B,e_A,e_B) & \implies & \eStartH{a}{b}\ \land\ \eStartH{b}{a}\\
		& & \land\ \mathit{userKey}(a,pk_A)\ \land\ \mathit{userKey}(b,pk_B) \\	
		& & \land\ \mathit{userEmail}(a,e_A)\ \land\ \mathit{userEmail}(b,e_B)
	\end{eqnarray*}
	In our model the email address is abstracted as the identity itself, since we consider the case of one account per user. Therefore, in the verification the \emph{userEmail} predicates are disregarded. We include them here for completeness.
	
	\paragraph{}
	\textbf{Trust-by-Handshake.}
	$${\displaystyle receiveGreen(b,a,m) \Longrightarrow receiverTrustsS(b,a)}$$
	This formula matches exactly the definition of Property \ref{p:trustByH}.
	
	\paragraph{}
	\textbf{Privacy-from-Trusted.}  For a message $z$,
	\begin{eqnarray*}
		\big(receiveGreen(b,a,z) &  \implies & \mathit{sendGreen}(a,b,z)\ \land\ z =aenc(m, pk_B) \\
		& & \land\ \mathit{\mathit{userKey}}(b,pk_B)\big)\  \land \\
		\big(decryptionFails(b,a,m) & \implies & \neg\,\mathit{sendGreen}( a,b,m)\big)
	\end{eqnarray*} 
	This formula is the conjunction of two correspondence assertions. The first one expresses Property \ref{p:pFt}; the second correspondence enforces the first by saying that it cannot be otherwise, i.e., when $b$ receives a message $m$ from $a$ which for any reason cannot be decrypted---e.g. $m$ is not encrypted---, then $a$ did not send $m$ to $b$.
	
	\paragraph{}
	\textbf{Integrity-from-Trusted.} For a message $z$ and a secret key $sk_A$
	\begin{eqnarray*}
		\big(receiveGreen(b,a,z) &  \implies & \mathit{sendGreen}(a,b,z)\  \land z =aenc(sign(m,sk_A), kb) \\
		& & \land\ \mathit{userKey}(a,sk_A) \big) \land \\
		\big(\mathit{signVerifFails}(b,a,m) & \implies & \neg\,\mathit{sendGreen}(a,b,m)\big)
	\end{eqnarray*} 
		%	$$signVerifFails(b,a,m) \implies \neg\, \mathit{sendGreen}(a,b,m)$$
	Analogous to the previous formula, in this one we express Property \ref{p:iFt} and reinforce it by proving that whenever the verification of the signature fails in message $m$, then $a$ did not send $m$. 
		
	\paragraph{}
	\textbf{MITM-detection.}
		\begin{eqnarray*}
			\mathit{endHandshakeUnsucc}(a,b,ka,kb) & \implies & (\mathit{userKey}(a,pk_A)\ \land\ pk_A\neq ka)\ \lor \\
			& & ( \mathit{userKey}(b,pk_B)\ \land\ pk_B\neq kb )
	\end{eqnarray*} 
	This formula matches exactly the definition of Property \ref{p:mitm}.

	\paragraph{}
	\textbf{Confidentiality.} 
	\emph{attacker} is a built in predicate in ProVerif, which evaluates to TRUE if by applying the derivation rules to the knowledge of the adversary, there exists a derivation that results in \emph{mssg}. Therefore, the protocol achieves confidentiality if $$\neg\,\mathit{attacker(mssg)}$$
}

\subsection{Verification Results and Analysis}
\label{subs:results}
In order to determine whether or not the protocol satisfies the specified security properties we use ProVerif \cite{proVerif}, an automatic symbolic cryptographic protocol verifier. 
We executed the verification\footnote{https://www.dropbox.com/s/ste22xe2zfj9bnt/fullPepProtocol.pv?dl=0} with ProVerif 2.0
on a standard PC (Intel i7 2.7GHz, 8GB RAM). The response time was immediate.

We analyzed three different models: of the key distribution protocol, of the trust establishment protocol and of the key distribution followed by the trust establishment (the \pep protocol).

For the key distribution protocol, the results confirmed its vulnerability to MITM attacks. 
The weakness resides in the exchange of public keys via a channel where \eve has complete access. An attack proceeds as follows:
\eve can intercept the initial message from \al to \bob and send him a new message attaching her own public key, $pk_E$, instead of \al's one. \pepb will then link \eve's key with \al's email in step (7) of Fig. \ref{fig:keyMngP}, i.e., $\mathit{storeId(idA_B,email_A,pk_E)}$. 
When \bob replies, the message in step (10) is encrypted with $pk_E$, and thus \eve can intercept it again and decrypt it with her secret key, therefore obtaining $pk_B$ attached.
From this point, \eve can send encrypted emails to \bob using \al's email address and she will be able to intercept and decrypt the responses sent by \bob. In an analogous way, \eve can have \al linking \eve's public key to \bob's identity, by sending her $pk_E$ encrypted with $pk_A$ obtained by intercepting the first message.

Regarding the trust establishment protocol, encryption and authentication hold since the trustwords comparison never mismatches due to the assumptions of the peer devices being secure and of a previous key distribution successfully executed.

\textbf{The subsequent analysis of  the \pep protocol determined that the six properties} (full agreement, trust-by-handshake, privacy-from-trusted, integrity-from-trusted, MITM-detection and confidentiality) \textbf{are satisfied}.

Regarding unsuccessful handshakes, even if \al has the correct public key of \bob, the handshake will fail if \bob has a key of \al that does not correspond to her. Both partners will mistrust each other because the communication with those keys is threatened, however, once a peer is mistrusted, by \pep design such a privacy rating can not be reverted. 
This might be an issue, for instance if in the future \al and \bob meet in person and exchange their public keys; they can then perform the handshake and \bob would be able to trust \al, but \al would not be able to trust \bob in her device.
In this case though, \eve misleading \al to mistrust the intended partner is closer to a Denial of Service (DoS) attack but does not represent a threat to privacy.

\textbf{We conclude that the execution of the \pep protocol fulfills the claimed security goals}, i.e., after a successful handshake there is no undetectable way for \eve to modify the exchanges between \al and \bob, given that every message between them is always sent encrypted and signed with the corresponding keys. As a consequence, the privacy, authentication and integrity of the messages is preserved. Also, entity authentication is achieved by the \pep trust establishment protocol.
These results depend on the assumptions of \pep residing in a secure environment, of a secure second channel for the \tws comparison and of \pep users owning a single instance of \pep with a single email account.

\subsection{Limitations}
\label{sec:discussion}
This analysis focuses solely on the technical specification of the key distribution and handshake protocols. Social attacks such as impersonation or phishing are however still possible; for instance \eve can create a fake email account related to \al's name and then use it to send \bob an email attaching \eve's public key and contact details. If \bob has never met \al, a handshake via \tws comparison with \eve would succeed given that both partners are indeed executing the protocol, but the human \bob thinks that he is interacting with the human \al.

The assumption of perfect cryptography implies that we consider the libraries implementing cryptographic operations to be correct. Implementation flaws in \pep and side-channel attacks are not considered either; however, 
we highlight the requirement for the software to ensure that the \tws database provided contains exactly the same data in all the distributions, to prevent introducing false mismatches during the \tws generation.

\section{Concluding Remarks}
\label{sec:discussion1}
We executed a symbolic \textbf{security analysis} of the specifications of \pep protocols for key distribution and authentication, \textbf{validating the exchange of authenticated end-to-end encrypted email between two \pep trusted peers}.
Here, we conclude by discussing some points that we have considered to extend this analysis in the future.

How humans behave when comparing \tws is not considered in this work; yet, incorrect input from users, such as mistrusting a trusted peer or vice-versa, might introduce security flaws.
These situations happen, for instance, when users verify only the first two words of the list or when they click the \tws confirmation button without comparing the \tws. 
A formal model of human errors in human-to-machine authentication protocols is proposed in \cite{basin2016modeling}; adapting such an approach to studying further the mentioned scenarios could give insights into how flaws introduced by users can be prevented. 
Understanding the causes and frequency of incorrect behavior requires a different kind of analysis mainly in the scope of usable security.

Another direction speculates whether solutions for automating security in \ac{im} can be applied in the context of email, as messaging protocols---e.g. Signal---achieve stronger security properties, such as forward secrecy.
The underlying reason preventing \pep from adopting similar approaches, hence upgrading security guarantees while depending less on the user, relies on the use of central servers; for instance, Signal uses a proprietary server as a deposit for ephemeral keys involved in the protocol. This is in opposition with the decentralized paradigm adopted in \pep's design, decision supported by the idea that ``the winner (i.e. the attacker) always takes all in centralized designs''.

\section*{Acknowledgments}
Authors were supported by the project pEp Security SA/SnT ``Protocols for Privacy Security Analysis''.

\bibliographystyle{splncs04}
\bibliography{biblio}

\end{document}